\documentclass[pre,amssymb,superscriptaddress,showpacs,floatfix,a4paper,twocolumn]{revtex4-1}

\usepackage{graphicx}% Include figure files
\usepackage{dcolumn}% Align table columns on decimal point
\usepackage{bm}% bold math
\usepackage{amsmath}
\usepackage{amssymb}
\usepackage{xcolor,cancel}

\begin{document}

\title{Micromagnetic simulation study of a disordered model for
one-dimensional granular perovskite manganite oxide nanostructures}

\author{P. Longone}
\affiliation{Departamento de F\'{\i}sica,  Universidad Nacional de
San Luis, INFAP, CONICET, Chacabuco 917, D5700BWS San Luis,
Argentina}
\author{F. Rom\'a}
\affiliation{Departamento de F\'{\i}sica,  Universidad Nacional de
San Luis, INFAP, CONICET, Chacabuco 917, D5700BWS San Luis,
Argentina}

\begin{abstract}

Chemical techniques are an efficient method to synthesize
one-dimensional perovskite manganite oxide nanostructures with a granular morphology, 
that is, formed by arrays of monodomain magnetic nanoparticles.
Integrating the stochastic Landau-Lifshitz-Gilbert equation, 
we simulate the dynamics of a simple disordered model for such materials 
that only takes into account the morphological characteristics of their nanograins.
We show that it is possible to describe reasonably well experimental hysteresis loops 
reported in the literature for single La$_{0.67}$Ca$_{0.33}$MnO$_3$ 
nanotubes and powders of these nanostructures, 
simulating small systems consisting of only 100 nanoparticles.

\end{abstract}

\maketitle

\section{Introduction}
At present many technological advances are strongly linked to both 
the synthesis of micrometric and submicrometric structures as well as the experimental 
and numerical studies of these materials \cite{Poole2003,Cao2004}.
Such is the case of the low-dimensional perovskite manganite oxide nanostructures,
nanoparticles, nanowires, nanotubes, nanofibers, and nanobelts,
which today are playing an important role in the development 
of new devices with applications in areas including microelectronic,
information storage and spintronic \cite{Fert1999,Handoko2010,Li2016}. 
Essentially, these technological innovations are possible because nanostructured materials 
have a large surface to volume ratio, implying that some of their physical properties 
can change significantly with respect to their bulk counterpart.  

One-dimensional perovskite manganite oxide nanostructures are synthesized by two methods \cite{Li2016}. 
With physical techniques, these magnetic systems are patterned from a bulk or a film
of the counterpart material by using lithography and etching. 
Instead, with different chemical approaches the nanostructures
are assembled from basic building blocks such as atoms or molecules.
In particular, a versatile and inexpensive chemical technique
uses porous sacrificial substrates of polycarbonate as templates
to produce ``granular'' nanowires and nanotubes, i.e.,
ultrafine and disordered assemblies of magnetic nanograins 
\cite{Levy2003,Leyva2004,Curiale2004,Curiale2007}. 
Since the characteristic diameter of these nanoparticles is very small,
they behave like single magnetic domains.
In addition, the existence of a magnetic dead layer (in each nanograin) \cite{Kaneyoshi1990},  
avoid exchange interactions among contiguous nanograins 
and therefore the dominating interactions are of dipolar type \cite{Curiale2007b}.  
  
The magnetic behavior of disordered granular nanowires and nanotubes 
has been little studied numerically in comparison to other
homogeneous and ordered one-dimensional nanostructures.
In order to explain the experimental resonance spectral data 
of La$_{0.67}$Sr$_{0.33}$MnO$_3$ (LSMO) manganite nanotubes,
Curiale {\it et al.} \cite{Curiale2008} have studied a model 
where each individual nanograin has an easy plane effective anisotropy.
The calculated resonance field agrees reasonably well with 
the experimental measurements made on samples of partially aligned nanotubes.  
By using a Monte Carlo Metropolis dynamics,
Cuchillo {\it et al.} \cite{Cuchillo2008} have calculated 
the hysteresis loops of a model of granular nanotube.  
Taking advantage of a geometric scaling property, 
the authors show that it is possible to simulate a small system 
to describe the behavior of a larger nanotube.  
Comparing with experimental data for again samples of LSMO nanotubes
they concluded, among other things, that the simulations neglecting dipole-dipole 
interaction never adjust to the experiment.

An in-depth study of such complex systems should require, in principle, 
to perform detailed simulations of a realistic model.
Since there is more than one aspect that contributes to the disorder, i.e.,
the nanograin size and shape are not uniform, the crystalline orientation is random,
the nanoparticles are arranged forming an amorphous structure, 
and even the number of magnetic moments is very huge,
a system that includes all these factors 
would be cumbersome to simulate by performing micromagnetic calculations 
based on the stochastic Landau-Lifshitz-Gilbert equation \cite{Brown1963}.
This differential equation gives the physically correct dynamical evolution
of a ferromagnetic system well below its Curie temperature.   
To overcome this problem, most studies have focused on using
Monte Carlo methods \cite{IglesiasChapter} which, in general, do not allow one to calculate 
the real dynamics of such magnetic systems. 
In addition, since the number of parameters of such a complex model 
could be very large, a subsequent comparison between the simulation results 
and the experimental data available in the literature
would not be useful to elucidate the contribution 
of each of these energy terms separately.

In this work, we show that it is not necessary to take into account
all these contributions to the disorder.
Performing micromagnetic calculations based on 
the stochastic Landau-Lifshitz-Gilbert equation, 
we simulate the real dynamic behavior of a simple disordered model:
a chain of nanograins with their uniaxial anisotropy axes oriented at random.
We show that, using typical values of parameters, 
it is necessary also to consider the dipolar interactions between nanograins  
to correctly describe a one-dimensional manganite nanostructure. 
Although the dynamics is not sensitive 
to the choice of the volume of nanograins,
we observe that it depends strongly on the anisotropy constant value. 
Assuming that the anisotropy constant is uniformly distributed,
we show that it is possible to fit very well the hysteresis loop
reported in the literature for two single La$_{0.67}$Ca$_{0.33}$MnO$_3$ (LCMO) nanotubes \cite{Dolz2008}.

The outline of the paper is as follows.  In Sec.\ref{ModelSimulation}
we introduce the model and the numerical micromagnetic scheme of calculation employed.   
Then, in Sec.\ref{Results} we present the results and discuss their implications.
Finally, Sec.\ref{Conclusions} is devoted to the summary and conclusions.  

\section{Model and simulation scheme \label{ModelSimulation}}

\subsection{One-dimensional disordered magnetic model}

In broad terms, granular manganite nanostructures (nanowires and nanotubes) 
are composed by at least tens to hundreds of thousands of nanograins 
whose typical sizes range between $4$ nm and $40$ nm, 
which are smaller than the critical diameter of a single magnetic domain \cite{Curiale2004,Curiale2007}.
Therefore, the magnetization of these nanoparticles can be represented 
by classical vectors of magnitude equal to the saturation magnetization. 
In some cases, given the material parameters and the non-spherical 
morphology of the nanograins which have an aspect ratio of approximately $1.5$,
it can be concluded that the shape anisotropy dominates over the crystalline one.   
Since in these nanostructured systems the nanoparticle assembly is disordered, 
this energy contribution can be taken into account by considering 
a uniaxial easy axis with random orientation for each nanograin. 
In addition, experimental Henkel plots \cite{Bertotti2008} show
that the dominating interactions between nanoparticles are of dipolar type \cite{Curiale2007b}.
This is due to the existence of a magnetic dead layer on the surface 
which avoids exchange interactions among nanograins.

In this context, we focus on studying a simple model in which only 
two types of disorder are taken into account. 
The first, and the most important one, 
is the random orientation of the uniaxial anisotropy axis of each nanoparticle
but, secondarily, we also include the possibility that the
anisotropy constant may vary locally.
Besides, we consider dipolar interactions between nanograins and,
to carry out the simulations, we use parameter values typically found in experiments.     

\begin{figure}[t]
\begin{center}
\includegraphics[width=8cm,clip=true]{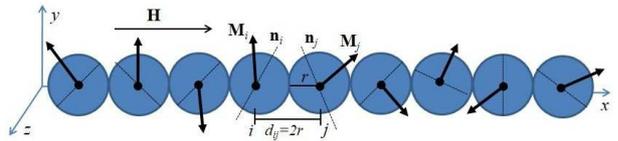}
\caption{Sketch of the model.  Each nanograin (sphere) of radius $r$ 
is separated from its first neighbors by a distance $d_{ij}=2r$ and 
has associated both a magnetization vector $\mathbf{M}_i$ and 
a random oriented uniaxial anisotropy axis $\mathbf{n}_i$.
In the figure the external field $\mathbf{H}$ is applied along the $x$ axis. }
\label{figure1}
\end{center}
\end{figure}

The energy per unit volume of the model is given by
\begin{eqnarray}
U&=& -\mu_0 \ \mathbf{H} \cdot \sum_{i=1}^N \mathbf{M}_i 
-\frac{1}{M_s^2}\sum_{i=1}^N K_i (\mathbf{M}_i \cdot  \mathbf{n}_i)^2  \nonumber \\
&& -\frac{\mu_0 V}{4 \pi} \sum_{i<j} \bigg[ \frac{ 3(\mathbf{M}_i \cdot \mathbf{e}_{ij})
(\mathbf{M}_j \cdot \mathbf{e}_{ij})- (\mathbf{M}_i \cdot \mathbf{M}_j)}{d_{ij}^3} \bigg]. 
\label{energy}
\end{eqnarray}
The first term is the Zeeman interaction,
the second represents the anisotropy energy,
and the last term is the dipolar coupling between nanograins. 
Here, $\mu_0$ is the vacuum permeability constant and 
$N$ is the number of nanograins of volume $V$, which are equally 
spaced and aligned along the $x$ axis; see Fig.~\ref{figure1}.
$\mathbf{M}_i$ is the magnetization at site $i$ whose magnitude is $M_s$, the saturation magnetization,
$\mathbf{H}$ is the external magnetic field, 
$\mathbf{n}_i$ is the uniaxial anisotropy axis vector and $K_i$ the corresponding constant,
and $\mathbf{e}_{ij}$ is a unit vector pointing from the site $i$ to the site $j$. 
Since in our model we have partially taken into account 
the structural disorder of granular one-dimensional perovskite manganite oxide nanostructures 
(only the random orientation of $\mathbf{n}_i$ and the possibility 
of having different local values of the anisotropy constant), 
we note that in Fig.~\ref{figure1} the nanograins are represented
by spheres of equal size and radius $r$, separated by a distance $d_{ij}=d=2r$.
Notwithstanding this simplification, the uniaxial anisotropy 
could have a magnetocrystalline or shape origin. 

\subsection{Micromagnetic simulation scheme}

We describe the dynamic time evolution of the classical magnetic moments 	
by the stochastic Landau-Lifshitz-Gilbert (sLLG) equation introduced by Brown \cite{Brown1963}
which, in the Landau formulation of dissipation \cite{Landau1935}, reads
\begin{eqnarray}
\frac{d\mathbf{M}_i}{dt}&=&-\frac{\gamma_0}{1+\gamma_0^2 \eta^2} \mathbf{M}_i \nonumber \\
&& \times \left[\mathbf{H}_i+\mathbf{W}_i+ \frac{\gamma_0 \eta}{M_s} \mathbf{M}_i
\times (\mathbf{H}_i+\mathbf{W}_i)\right],
\label{sLLG}
\end{eqnarray}
where $t$ is the time (in seconds) and $\gamma_0\equiv\gamma \mu_0=2.2128 \times 10^5$ m/(As),
with $\gamma$ being the gyromagnetic ratio.
In our convention $\gamma>0$ and is given by $\gamma=\mu_B g/\hbar$,
with $\mu_B$ the Bohr's magneton, $g$ the Lande's $g$-factor, 
and $\hbar$ the reduced Planck's constant.
$\mathbf{H}_i$ is the local effective field acting at each site given by 
$\mathbf{H}_i=-\mu_0^{-1} \partial U / \partial \mathbf{M}_i$ which, for our model, reads 
\begin{eqnarray}
\mathbf{H}_i &=& \mathbf{H} + \frac{2 K_i}{M_s^2 \mu_0} (\mathbf{M}_i \cdot  \mathbf{n}_i)  \mathbf{n}_i \nonumber \\
&& +\frac{V}{4 \pi} \sum_{j \ne i} \bigg[ \frac{ 3(\mathbf{M}_j \cdot \mathbf{e}_{ij}) \mathbf{e}_{ij}
- \mathbf{M}_j}{d_{ij}^3} \bigg]. 
\label{EffectiveField}
\end{eqnarray}
Thermal effects are introduced by random fields $\mathbf{W}_i$  
which are assumed to be Gaussian distributed with average 
\begin{equation} 
\langle W_{i,k}(t) \rangle_{\mathbf W} = 0 
\label{average}
\end{equation}
and correlations \cite{Brown1963}
\begin{equation} 
\langle W_{i,k}(t) W_{i,l}(t')  \rangle_{\mathbf W} = 2 D \ \delta_{kl} \ \delta(t-t'),
\label{correlations}
\end{equation}
for all $k,l=x,y,z$ components. The parameter $D$ is chosen as
\begin{equation}
D = \frac{\eta k_B T}{M_s V \mu_0} \label{paramaterD},
\end{equation}
so that the sLLG equation takes the magnetization to equilibrium at temperature $T$ \cite{Aron2014}.
$\eta$ is the phenomenological damping constant and $k_B$ is the Boltzmann's constant. 

As usual, we introduce the adimensional time $\tau=\gamma_0 M_s t$ 
and the adimensional damping constant $\eta_0=\eta \gamma_0$, 
and we normalize all fields by $M_s$, 
$\mathbf{m}_i=\mathbf{M}_i/M_s$, $\mathbf{h}=\mathbf{H}/M_s$, 
$\mathbf{h}_i=\mathbf{H}_i/M_s$, and $\mathbf{w}_i=\mathbf{W}_i/M_s$, to write the Eqs. 
(\ref{sLLG}), (\ref{EffectiveField}), (\ref{average}), and (\ref{correlations}) as,
respectively,
\begin{eqnarray}
\frac{d\mathbf{m}_i}{d\tau}&=&-\frac{1}{1+\eta_0^2} \mathbf{m}_i \nonumber \\
&& \times \left[\mathbf{h}_i+\mathbf{w}_i+ \eta_0 \ \mathbf{m}_i
\times (\mathbf{h}_i+\mathbf{w}_i)\right],
\label{sLLG_normal}
\end{eqnarray}
\begin{eqnarray}
\mathbf{h}_i &=& \mathbf{h} + \frac{2 K_i}{M_s^2 \mu_0} (\mathbf{m}_i \cdot  \mathbf{n}_i)  \mathbf{n}_i \nonumber \\
&& +\frac{V}{4 \pi} \sum_{j \ne i} \bigg[ \frac{ 3(\mathbf{m}_j \cdot \mathbf{e}_{ij}) \mathbf{e}_{ij}
- \mathbf{m}_j}{d_{ij}^3} \bigg], 
\label{EffectiveField_normal}
\end{eqnarray}
\begin{equation} 
\langle w_{i,k}(\tau) \rangle_{\mathbf w} = 0, 
\label{average_normal}
\end{equation}
and 
\begin{equation} 
\langle w_{i,k}(\tau) w_{i,l}(\tau')  \rangle_{\mathbf w} = \frac{2 D \gamma_0}{M_s} \ \delta_{kl} \  \delta(\tau-\tau').
\label{correlations_normal}
\end{equation}

The sLLG is a Markovian first-order stochastic differential equation 
characterized by having a multiplicative thermal white noise coupled to magnetization. 
This implies that to completely define the dynamics,
it is required to specify a ``prescription'' for the way in
which the noise acts at a microscopic time level.
To preserve the magnetization module
(for a ferromagnetic system well below the Curie temperature
one is interested in describing the evolution of the magnetization keeping its modulus fixed)
it is necessary to use the Stratonovich mid-point prescription, stochastic calculus 
\cite{Aron2014,Bertotti2009,Gardiner1997}.
From a practical point of view, the normalized sLLG equation (\ref{sLLG_normal}) 
can be easily integrated using the Heun method which
converges to the solution interpreted in the sense of the
Stratonovich explicit discretization scheme \cite{Rumelin1982,GarciaPalacios1998}.
In Cartesian coordinates, this simulation scheme require 
the explicit magnetization normalization after every time step $\Delta t$ \cite{Martinez2004,Cimrak2007}.
We use an adimensional time step of $\Delta \tau = 0.01$, 
equivalent to $\Delta t = \Delta \tau /(\gamma_0 M_s) \approx 0.08$ ps,
which is sufficiently small to ensure convergence from further reductions in $\Delta \tau$.
In addition, the simulations were performed choosing $\eta_0 = 0.01$.

In all cases we calculate hysteresis loops starting from the saturation state
and sweeping the external field at different rates $R$.
The average value of the total normalized magnetization
\begin{equation}
\mathbf{m}=\frac{1}{N} \Bigg\langle \sum_{i=1}^N \mathbf{m}_i \Bigg\rangle,
\end{equation}     
and their components, $m_x$, $m_y$, and $m_z$, 
are calculated at low temperatures well below the Curie critical temperature
(for some perovskite manganite oxide nanostructures this critical temperature 
is of the order of $300$ K \cite{Curiale2007}). 
For simplicity and without causing confusion, in some cases 
$\langle ...\rangle$ represents only an average 
over many cycles for a single nanostructure (over at least $10^2$ cycles),  
but in other situations also includes an additional average
over disorder (over about $10^2$ different nanostructures).

\section{Results and discussion \label{Results}}

\begin{figure}[t!]
\begin{center}
\includegraphics[width=6cm,clip=true]{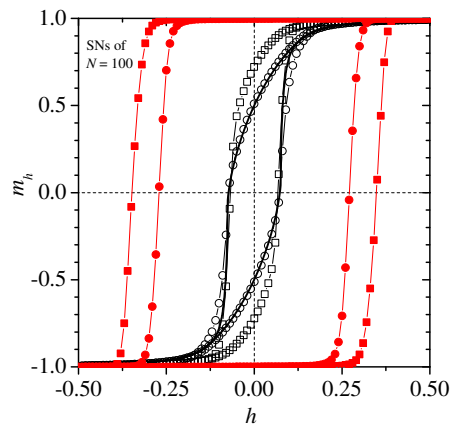}
\caption{Hysteresis loops calculated for SNs of size $N=100$, 
with their anisotropy axis vectors oriented at random (black open symbols) 
and aligned along the $x$ axis (red closed symbols).
Squares and circles stand for curves with and without dipolar interactions, respectively. 
The continuous thick line corresponds to the hysteresis loop for the RSW model. } 
\label{figure2}
\end{center}
\end{figure}

We carried out the micromagnetic simulations for five different sweep rates: 
$R_1=2.3 \times 10^{14}$ A/(m s), 
$R_2=2.3 \times 10^{13}$ A/(m s), $R_3=2.3 \times 10^{12}$ A/(m s),
$R_4=2.3 \times 10^{11}$ A/(m s), and $R_5=2.3 \times 10^{10}$ A/(m s).
These values of $R$ are relatively high, since integrating the sLLG equation 
requires one to use very short-time steps and then it is possible only to calculate
high-frequency hysteresis loops.  However, as we will discuss later,
the hysteresis loops calculated for the lowest values of $R$ 
should not be very different from those that 
would be obtained at typically experimental low sweep rates.  
In general we apply the external field along the $x$ axis but, 
when this is not the case, 
the orientation of $\mathbf{h}$ will be explicitly indicated. 

We investigate the effects of disorder following two strategies. 
First, we study the hysteresis loops of a single nanostructure, i.e., 
a single disorder realization as sketched in Fig.~\ref{figure1}, 
analyzing the influence of each energy term in Eq.~(\ref{energy}). 
Next, we analyze these curves averaged over disorder 
to mimic the magnetic behavior of ensembles (powder samples) of these nanostructures.
In both cases we consider the possibility of having 
the anisotropy axis oriented at random, 
but we use a single value of $K_i=K$. 
We choose $T=25$ K and we use typical values of material parameters, 
namely, $d=22.8$ nm ($V=6.2 \times 10^{-24}$ m$^3$), $M_s=5.8 \times 10^5$ A/m, 
and $K=3.2 \times 10^4$ J/m$^3$.
After this general study we focus on simulating a particular system, LCMO nanotubes.
We will see that, in addition to the random orientation of $\mathbf{n}_i$,
it is also necessary to take into account the local variations 
of the anisotropy constant to fit well the experimental data.      
 
\subsection{Single nanostructures}
 
We simulate single nanostructure (SNs) of different sizes at the intermediate sweep rate $R_3$.
Figure~\ref{figure2} shows the hysteresis loop (black open squares) 
for a typical ``disordered nanostructure'' of size $N=100$.
Here, $h$ is the magnitude of the normalized external field
(in this case $h=h_x$, its $x$ component),
while $m_h$ is the projection of $\mathbf{m}$ onto the direction of $\mathbf{h}$
(in this case $m_h=m_x$, the $x$-component of the total normalized magnetization). 
We show also an equivalent curve (red closed squares) calculated 
for a ``nondisordered nanostructure'' of the same size, or more precisely, 
a chain of $N$ nanograins with their anisotropy axis vectors aligned along the $x$ axis.
As expected, in the last loop the values of the coercive field, $h_c$, 
and the remanent magnetization, $m_r$, increase with respect to the disordered case. 
Besides, for both types of nanostructures, we present in Fig.~\ref{figure2}
the hysteresis loops (open and closed circles) calculated by removing 
the dipolar term of the effective field.
We see clearly that such elimination produces significant changes in these curves,
showing that none of the interactions (anisotropy and dipolar interactions) 
dominates the dynamics by itself.  
In fact, from Eq.~(\ref{EffectiveField_normal}) we can easily verify that,
when comparing maximum values, the second and third term of the effective field
are of the same order of magnitude.  
Therefore, we conclude that both the anisotropy and the dipolar interactions, 
as well as the structural disorder, play a preponderant role
in determining the dynamical behavior of these manganite nanostructures  
(remember that we are using typical values of material parameters 
found in the literature for this system).

\begin{figure}[t!]
\begin{center}
\includegraphics[width=6cm,clip=true]{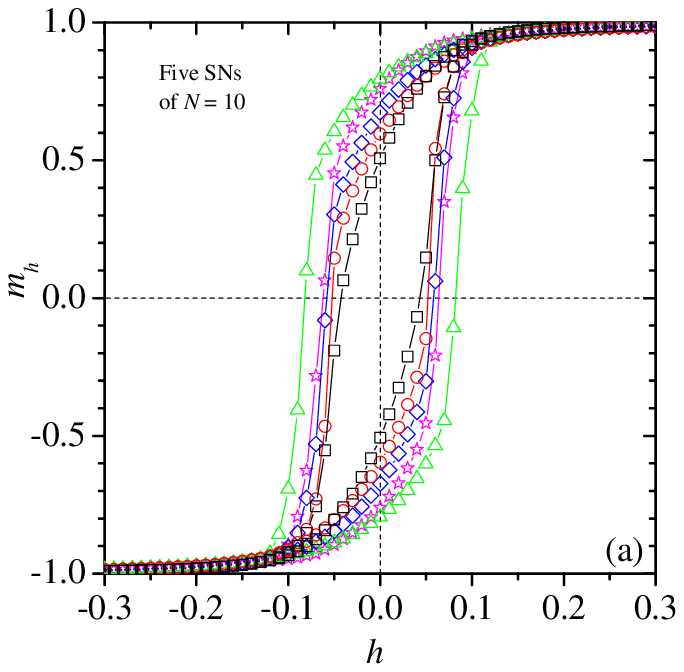}
\includegraphics[width=6cm,clip=true]{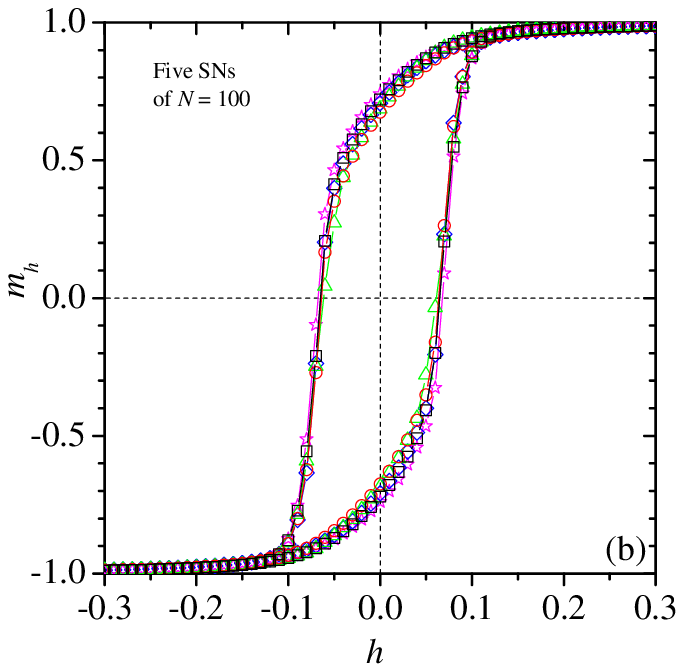}
\caption{Hysteresis loops calculated 
for five different SNs of size (a) $N=10$ and (b) $N=100$.   } 
\label{figure3}
\end{center}
\end{figure}

\begin{figure}[b!]
\begin{center}
\includegraphics[width=6cm,clip=true]{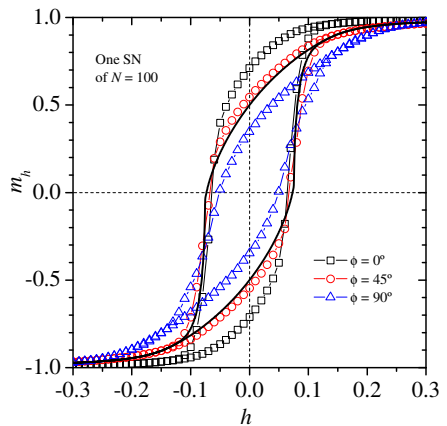}
\caption{Hysteresis loops for one SN of size $N=100$,
calculated for three different values of $\phi$ as indicated.
The continuous thick line corresponds to the hysteresis loop for the RSW model.} 
\label{figure4}
\end{center}
\end{figure}

It is instructive to make a comparison with an assembly of noninteracting nanograins 
with their easy axes randomly oriented in space at temperature $T=0$, 
which is none other than the random Stoner-Wohlfarth model (RSW) \cite{Stoner1948,Cullity}.
In Fig.~\ref{figure2} we show that the hysteresis loop for this system (continuous thick line)
agrees pretty well with the simulation results of our model without dipolar interactions (black open circles).
This is so due to the simulations take place in both,
the low-temperature regime (the ratio $V \Delta U_0/k_B T \ll 1$, 
where $V \Delta U_0$ is the height of a typical energy barrier)
and the low-dynamic regime (the sweep rate $R_3$ is low enough 
so that the numerical results are expected to become close 
to those obtained from static calculations; see below).
As usual, the remanent magnetization for the RSW model is $m_r=0.5$ but,
due to the normalization chosen,
the coercive field is $h_c \cong 0.958 K / \mu_0 M_s^2=0.07252$ \cite{Bertotti2008}.  

The number of nanograins is another important factor.  
Figures~\ref{figure3}(a) and 3(b) show the hysteresis loops 
for two sets, each with five different SNs, 
of size $N=10$ and $N=100$, respectively.
When the size increases logically the dispersion in the values of $h_c$ and $m_r$ decreases.
Through additional simulations (up to $N=10^3$), we have corroborated that
SNs with a relatively small size (e. g., with $N \ge 100$)
should represent qualitatively well the dynamical behavior of much larger structures. 
Therefore, most of our calculations have been performed for $N=100$.

\begin{figure}[t!]
\begin{center}
\includegraphics[width=6cm,clip=true]{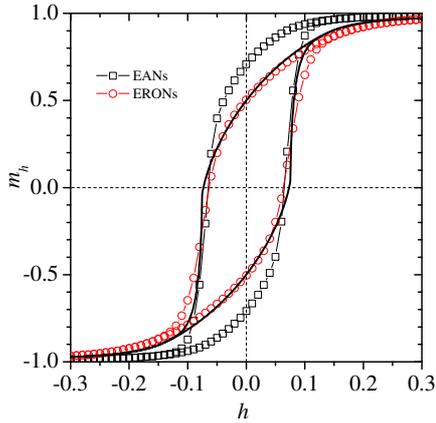}
\caption{Hysteresis loops calculated 
for (a) an EANs and (b) an ERONs of sizes $N=100$.
The continuous thick line corresponds to the hysteresis loop for the RSW model.} 
\label{figure5}
\end{center}
\end{figure}

On the other hand, when applying the external field 
along a direction other than the easy axis of magnetization,
in particular at an angle $\phi$ with respect to the $x$ axis,
the hysteresis loops change significantly \cite{note1}.
Figure~\ref{figure4} shows the curves for a SN of size $N=100$,
for $\phi=0^{\circ}$, $\phi=45^{\circ}$, and $\phi=90^{\circ}$
(note that now $m_h=m_x$ only when $\phi=0^{\circ}$). 
Whereas $m_r$ decreases appreciably, the value of $h_c$ changes 
very little as the angle $\phi$ increases.
The reasons are as follows.  
In the saturation state for $\phi=0^{\circ}$,   
the dipolar contribution to the effective field Eq.~(\ref{EffectiveField_normal})
points in the same direction as the external field,
while for $\phi=90^{\circ}$ these fields are opposed to each other.
Therefore, when $h$ becomes zero the demagnetization process 
is more effective in the latter case and the corresponding remanent magnetization decreases.
An intermediate situation occurs for $\phi=45^{\circ}$.
In this case, close to the saturation state the dipolar field is nearly transversal to $h$ \cite{note2} and,
although the three components of magnetization are not independent of each other,
its effect on $m_h$ is very small, affecting to a large extent
the two components of magnetization perpendicular to this direction.  
As the dynamic is governed mainly by the external and anisotropy fields,
the hysteresis loop for $\phi=45^{\circ}$ is very close to that of the RSW model (see Fig.~\ref{figure4}). 
In addition, the coercive field values change very little with $\phi$,
since at zero magnetization the dipolar field is very small
and the anisotropy term dominates (which does not depend on how the nanograins are arranged).
As shown below, this angular dependence of the hysteresis loop 
allows us to explain what is observed 
in ensembles of randomly oriented nanostructures.

\subsection{Ensembles of nanostructures}

\begin{figure}[t]
\begin{center}
\includegraphics[width=6cm,clip=true]{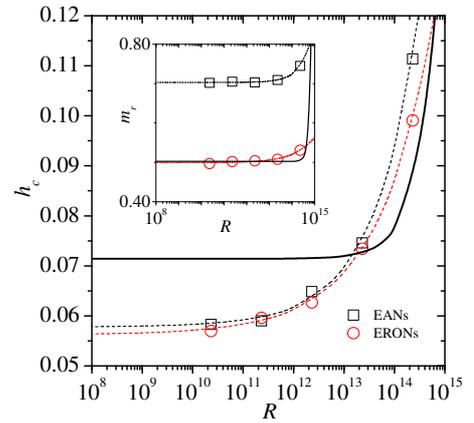}
\caption{Coercive field versus $R$ for both an EANs and an ERONs.
Inset shows the same information for the remanent magnetization.  
Dashed and dotted curves are fits to Eq.~(\ref{fit}).
The continuous thick line both in the main panel and inset are,
respectively, the curves of $h_c$ and $m_r$ as functions of $R$ 
for an assembly of noninteracting nanograins with their easy axes randomly oriented in space.} 
\label{figure6}
\end{center}
\end{figure}

Experimentally, it is possible to prepare samples of partially aligned 
one-dimensional nanostructures \cite{Curiale2007}: a powder of randomly 
oriented nanostructures is diluted in ethyl alcohol, next the solution is deposited 
on a glass substrate and is placed on a uniform magnetic field, 
and finally the solvent is evaporated.  To mimic (approximately) this configuration,
here we study an ensemble of (non-interacting equal-sized) 
aligned nanostructures (EANs).  We simply average over disorder 
the hysteresis loops of SNs of the same size.  Since in this calculation 
we do not consider the interactions between nanostructures,
we expect our results to be only qualitatively significant 
to understand the magnetic behavior of this type of sample.  
   
Figure~\ref{figure5} shows the hysteresis loop for an EANs 
of size $N=100$ (black open squares) calculated at the sweep rate $R_3$,
where the external field is applied along the $x$ axis.
As there are no interactions between nanostructures, 
we note that this curve is essentially the same as the corresponding 
hysteresis loop in Fig.~\ref{figure1} for SNs
(since the convergence with respect to $N$ is very fast,
the curves for a SN and for an EANs of the same size
are very close to each other). 
On the other hand, in Fig.~\ref{figure5} we also show the hysteresis
loop for an ensemble of (non-interacting equal-sized) 
randomly oriented nanostructures (ERONs),
a system that should resemble approximately the magnetic behavior 
of a typical powder sample (of unaligned nanostructures). 
In addition to considering the structural disorder of each SN,
here we average also over the different directions of their main axes 
assuming that these are randomly oriented.
We note that the curves have similar coercive fields,
but the remanent magnetization is significantly smaller for an ERONs than an EANs.
This behavior is expected when we consider how the curves depend on 
the orientation of the applied external field for one SN; see Fig.~\ref{figure4}.
  
Interestingly, Fig.~\ref{figure5} also shows that the shape of the hysteresis loop 
for an ERONs is very similar to that of the RSW model.
This is logically a consequence of a compensation effect since, according to Fig.~\ref{figure4}, 
the curve for $\phi=45^{\circ}$ (which is also close to that of the RSW model, see above) 
represents an intermediate case between the $\phi=0^{\circ}$ and $\phi=90^{\circ}$ ones, 
and therefore an average over different directions should lead to a hysteresis loop very close to it. 

In Fig.~\ref{figure6} we show how depend the coercive field with the rate $R$
for both an EANs and an ERONs, and in the inset we present the same information 
for the remanent magnetization.  
Approximately at $R \approx R_3$, these curves exhibit a crossover from
a high dynamic regime (steep increase of coercive field with $R$) 
to a low dynamic regime (a slow variation of coercive field with $R$) \cite{Moore2004}.
The crossover occurs when the simulation timescale approaches the timescale 
of the intrinsic switching process, typically on the order of nanoseconds for 
the coherent rotation of the magnetization of single-domain particles.

Dashed and dotted curves in Fig.~\ref{figure6} are fits to the following equation:
\begin{equation}
X(R)=X(0)+a R^b, \label{fit}  
\end{equation}
where $X(R)$ [$X(0)$] stand for both $h_c(R)$ [$h_c(0)$] and $m_r(R)$ [$m_r(0)$],
and $a$ and $b$ are additional fit parameters.
In the limit of zero or very low sweep rates (the relevant frequencies for the experiments), 
we obtain $h_c(0) = 0.058(1)$ and $m_r(0) = 0.703(1)$ for an EANs
and $h_c(0) = 0.056(1)$ and $m_r(0) = 0.500(2)$ for an ERONs.
It is very important to note that the coercive fields calculated for 
$R_3$, $R_4$, and $R_5$ are, respectively, only 
$12\%$, $2\%$, and $1\%$ larger than limit values.
This means that our simulation results for $R_3$ 
can be compared at least qualitatively with experiments,
while for $R_4$ and $R_5$ we can even make a quantitative comparison. 
The same conclusion is valid for the remanent magnetization. 

We make again a comparison with a disordered assembly of nanograins without dipolar interactions. 
In Fig.~\ref{figure6} we show the curves of $h_c$ and $m_r$ as functions of $R$ 
for this system obtained in our simulations.  
The coercive field and the remanent magnetization (inset) tends, respectively, 
to $h_c(0) = 0.071(2)$ and $m_r(0) = 0.503(3)$ values that, as expected, 
agree well with that calculated for the RSW model ($h_c \cong0.0725$ and $m_r=0.5$),
and also in particular the remanence value is very close to the corresponding one 
for an ERONs in the limit of $R \to 0$.
On the other hand, we see that the coercivity for an EANs and an ERONs
are lower than the one obtained for non-interacting nanograins.
This behavior is in line with recent Monte Carlo simulations of our one-dimensional model.
In Ref.~\cite{IglesiasChapter} (see Chap. 3) it was observed that, at low temperatures, 
$h_c$ decreases with increase of the intensity of the dipolar interaction.  
Clearly this effect is produced by the disorder because, as we have seen before in Fig.~\ref{figure2},
the coercivity increases when the interactions are turned on 
if all the anisotropy axes are aligned along the $x$ axis.

\subsection{Manganite nanotubes}

In previous subsections, we have analyzed the hysteresis loops 
of SNs and different ensembles using typical parameters of real systems.   
Now, we will try to go one step further by comparing
our simulation results with experimentally obtained data. 
 
In Ref.~\cite{Dolz2008}, the hysteresis loop for two single 
(magnetically isolated) LCMO nanotubes was measured using 
a silicon micromechanical torsional oscillator working in its resonant mode. 
Also, a commercial superconducting quantum interference device 
was used to measure the same curve for a powder of randomly oriented LCMO nanotubes.
A simple comparison between our Fig.~\ref{figure5} (calculated using typical parameter values) 
and Fig. 4 of Ref.~\cite{Dolz2008} shows that, at least on a qualitative level,
our simple model reproduces quite well the experimental data.
In particular, in the simulations we observed that the coercive fields 
for an EANs (we remember that the hysteresis loop 
for an EANs is essentially the same as for a SN) and for an ERONs  
are almost the same, while for the experimental curves both values     
are very close to each other. 
Besides, for both simulations and experiments, the remanent magnetization
value for an EANs is appreciably higher than for an ERONs.

Nevertheless, it is still possible to go further and to make a quantitative comparison.
For that we must use a specific parameter set for this material. 
First, we focus on the experiments for single LCMO nanotubes 
for which the saturation magnetization measured 
at $T=14$ K was $M_s=3 \times 10^5$ A/m \cite{Dolz2008}.
It is important to highlight that
we restrict the analysis to the range $|H| \le 1.8 \times 10^5$ A/m,
within which the ferromagnetic core of each nanograin saturates,
since for larger field intensities the magnetization continues to increase due 
to the contribution of the magnetic dead layer \cite{Curiale2009},
a phenomenon that our model is not able to reproduce.
We can keep $M_s$ fixed but not 
the diameter $d$ and the anisotropy constant $K_i$ of each nanograin. 
According to a morphological study \cite{Curiale2007},
a LCMO nanotube is formed by nanoparticles whose diameter distribution ranges 
from $10$ nm to $45$ nm, and has a maximum at $25$ nm.
On the other hand, as we mentioned earlier, given the non-spherical
morphology of the nanograins it is clear that the shape anisotropy,
which for a prolate ellipsoid with a typical aspect ratio of $1.5$
corresponds to an uniaxial anisotropy constant of $K=3.2 \times 10^4$ J/m$^3$ \cite{Cullity},
dominates over the crystalline one that, for this type of materials, 
is $K\approx 1 \times 10^3$ J/m$^3$ \cite{Suzuki1998}.  
In addition, all nanoparticles do not have the same aspect ratio 
(this distribution is unknown) but ferromagnetic-resonance experiments performed
in LSMO nanotubes (which has a similar nanostructure to the LCMO nanotubes) 
confirm that the average anisotropy constant 
is approximately $K \approx 2.9 \times 10^4$ J/m$^3$ \cite{Curiale2007}.    

\begin{figure}[t!]
\begin{center}
\includegraphics[width=6cm,clip=true]{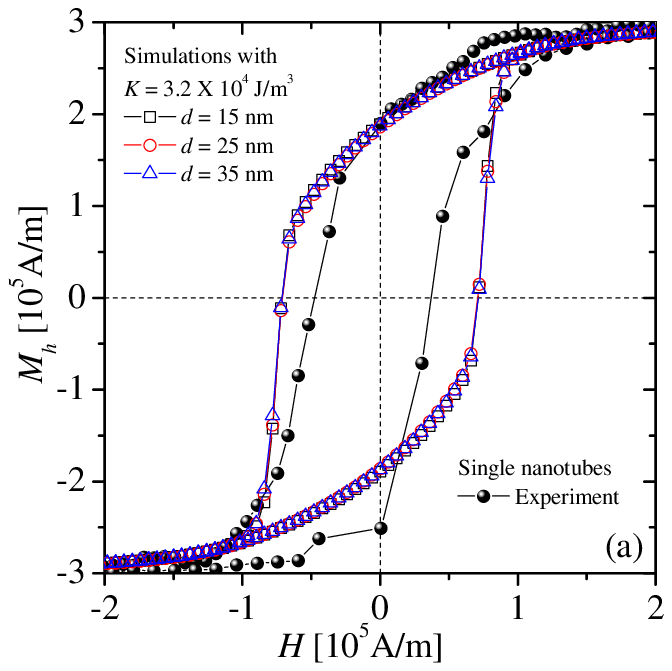}
\includegraphics[width=6cm,clip=true]{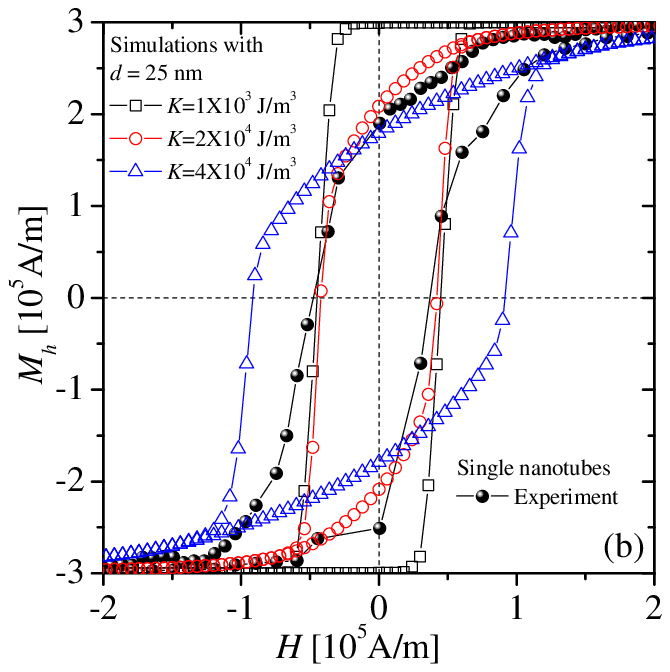}
\caption{Experimental hysteresis loop for two single 
magnetically isolated LCMO nanotubes (black spheres)\cite{Dolz2008}, 
along with simulation curves calculated for (a) different diameters $d$  
(with $K$ constant) and for (b) different values of $K$ (with $d$ constant),
as indicated.  } 
\label{figure7}
\end{center}
\end{figure}

In this context, we analyze how the hysteresis loops depend 
on the diameter of the nanograins and their anisotropy constants.
To achieve a good agreement with the experimental data, 
in what follows we carry out the simulations at the sweep rate $R_4$.
Figure~\ref{figure7}(a) shows the curve
obtained for single LCMO nanotubes \cite{Dolz2008}, 
compared with the corresponding one calculated by simulations of EANs
of size $N=100$, for three different values of diameter, $d=15$ nm, $25$ nm, and $35$ nm,
using the same value of $K=3.2 \times 10^4$ J/m$^3$.
Instead of normalized quantities, 
we now plot the magnetization $M_h$ ($M_h=m_h M_s$) versus the field $H$ in MKS units. 
Numerical hysteresis loops are very similar to each other
and do not fit well the experimental data.
This evident insensitivity to changes in the volume can be explained  
by noting first that the effective field Eq.~(\ref{EffectiveField_normal})
does not depend on $V$ (although the dipolar term is proportional to $V$,
also is inversely proportional to $d_{ij}^3 = d^3 = 6 V / \pi$).
Only the parameter $D$, Eq.~(\ref{paramaterD}), changes with the volume but,
since the temperature is sufficiently low, the dynamics of the system is not very affected
showing that the exact size of the nanograins is not relevant
(justifying our original assumption of having nanoparticles of equal volume).
We have corroborated that this is also the case when we choose different
values of the adimensional damping constant, $\eta_0=0.05$, $0.01$, and $0.005$.
 
Instead, the hysteresis loop shape is sensitive to variations of the anisotropy constant. 
As before, we show in Fig.~\ref{figure7} (b) the experimental curve 
obtained for single LCMO nanotubes, but now we compare it 
with simulations results of EANs for three different values of $K$ 
keeping constant the diameter in $d=25$ nm.
For $K=1 \times 10^3$ J/m$^3$, the magnitude of the crystalline anisotropy,
the numerical hysteresis loop shows a square like shape
that is far from fitting the experimental data.
Since the value of the anisotropy constant is very low, 
the dipolar field dominates the dynamical behavior. 
This should be the case if the nanograins were spherical
(which corresponds to an aspect ratio of one).
On the other hand, for $K=2 \times 10^4$ J/m$^3$ and $K=4 \times 10^4$ J/m$^3$,
the shape anisotropy constant of nanograins with aspect ratios 
of approximately $1.3$ and $1.7$, respectively,   
the curves display the characteristic $S$ shape 
of the experimental hysteresis loop.
Nevertheless, we see that it is not possible to use 
a single value of $K$ to correctly represent the 
real behavior of single nanotubes. 

\begin{figure}[t]
\begin{center}
\includegraphics[width=6cm,clip=true]{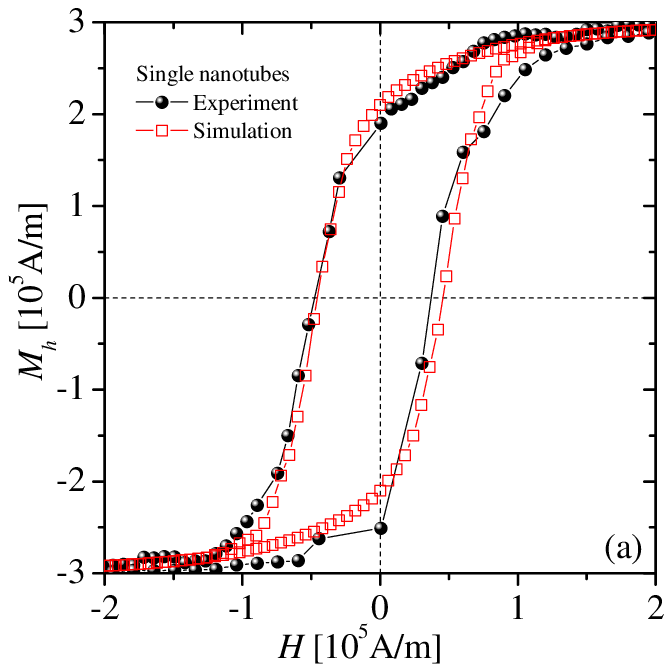}
\includegraphics[width=6cm,clip=true]{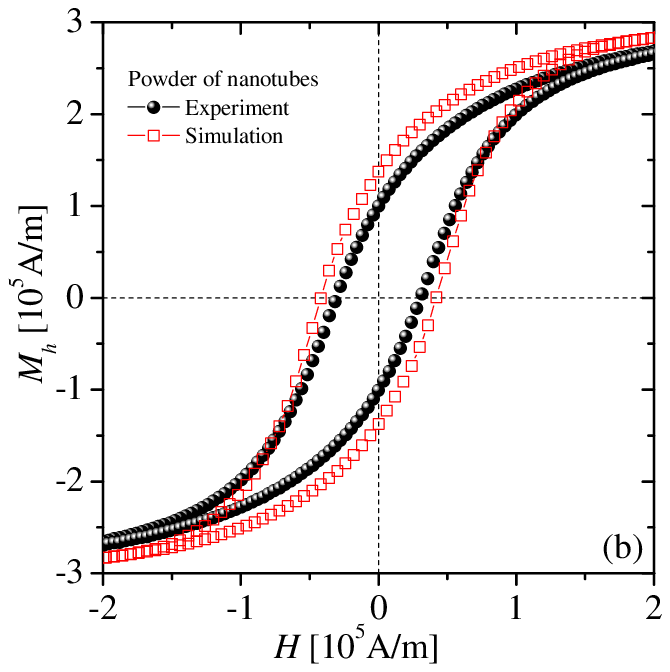}
\caption{Experimental hysteresis loops for 
(a) two single magnetically isolated LCMO nanotubes and for 
(b) a powder of this material \cite{Dolz2008},
along with simulation curves calculated using
the parameters $K_0=2.5 \times 10^4$ J/m$^3$ and $\Delta K=2.0 \times 10^4$ J/m$^3$.} 
\label{figure8}
\end{center}
\end{figure}

These results tell us that, in addition to the random orientation 
of the anisotropy axes, it is also necessary to consider 
the local variations of the anisotropy constant
in order to try to fit the experimental data.  
Since the distribution of this constant is unknown,
for simplicity we choose to work with a uniform distribution
of $K_i$ between $K_0-\Delta K$ and $K_0+\Delta K$,
with $K_0$ being the mean value of this constant. 
In Fig.~\ref{figure8}(a) we show how the numerical 
hysteresis loop, calculated for an EANs of size $N=100$ with nanograins of diameter $d=25$ nm,
matches very well with the experimental data for single LCMO nanotubes
if we choose $K_0=2.5 \times 10^4$ J/m$^3$ and $\Delta K=2.0 \times 10^4$ J/m$^3$.
We note that the asymmetry of the last curve  
is clearly due to experimental uncertainties 
which can arise in these kinds of delicate experiments.  
Approximately, the value of $K_0=2.5 \times 10^4$ J/m$^3$ 
corresponds to an aspect ratio of $1.37$,
while the extremes $K_0-\Delta K=5.0 \times 10^3$ J/m$^3$ and 
$K_0+\Delta K=4.5 \times 10^4$ J/m$^3$ correspond to,
respectively, aspect ratios of $1.06$ and $1.82$. 
This range of values of the anisotropy constant is consistent with the morphological 
characteristics observed experimentally \cite{Curiale2007}.

With the same set of parameters we can try to fit 
the experimental data for powders of randomly oriented LCMO nanotubes.
Figure~\ref{figure8}(b) shows a comparison between
the hysteresis loop measured in Ref.~\cite{Dolz2008} 
and the simulation result for an ERONs of size $N=100$.  
The shapes of the curves are very similar but,
being that in our calculation we have not considered 
the interactions between nanostructures,
as it was expected there are appreciable differences 
between simulation and experiments. 
Both $h_c$ and $m_r$ values for the powder sample 
are lower than the numerical ones. 
Possibly, this is due to the dipolar contributions to the effective field
produced by the nanograins that are around the nanotube 
and that do not belong to this nanostructure. 
These ``external'' nanoparticles are located outside the main axis
and, on average, at a distance greater than the typical diameter size $d$
which, in general terms, should produce a small decrease in the effective field in each site.

\section{Summary and conclusions \label{Conclusions}}

In this work, we have studied numerically the dynamic 
behavior of a simple disordered model for 
one-dimensional granular perovskite manganite oxide nanostructures.
The model includes only two types of disorder, 
both linked to the morphological characteristics of the nanograins: 
the random orientation of their uniaxial anisotropy axes $\mathbf{n}_i$ and 
the local variations of the anisotropy constant $K_i$.
The dynamics is simulated by integrating the stochastic Landau-Lifshitz-Gilbert equation,
using parameter values typically found in experiments.

First, we consider only the variations in $\mathbf{n}_i$ keeping constant $K_i=K$. 
Since the convergence with size is very fast, 
we show that small systems of 100 nanoparticles
are large enough to describe the dynamics correctly.
We analyze then the hysteresis loops for both SNs
and two different ensembles of noninteracting nanostructures, EANs and ERONs. 
Our simulations agree qualitatively well with the experimental results
reported in the literature for single LCMO 
nanotubes and powders of this nanomaterial.

To make a quantitative comparison with this experimental data,
we analyze the curves with respect to variations in both
the diameter of the nanograins and their anisotropy constants.
We conclude that only changes in $K$ significantly affect the shape
of the hysteresis loops. Choosing a uniform distribution for the 
local anisotropy constant $K_i$,
we show that it is possible to fit very well the hysteresis loop for single LCMO nanotubes.  
Finally, although in the ERONs we have neglected the interactions
between nanostructures, our simulations are close to the experimental 
data for powders of LCMO nanotubes.

In conclusion, we have shown that it is possible to describe reasonably 
well experimental hysteresis loops of granular manganite nanotubes,
simulating the true dynamical behavior of a simple model 
without taking into account further complex aspects (the tubular shape
of nanotubes) that also contribute to the disorder.     
 
\begin{acknowledgments}

We thank M. I. Dolz for providing us the experimental data of Ref.\cite{Dolz2008}.
This work was supported in part by CONICET under Project No. PIP 112-201301-00049-CO, 
by FONCyT under Project No. PICT-2013-0214, and by Universidad Nacional de San Luis 
under Project No. PROICO P-31216 (Argentina). 

\end{acknowledgments}


\begin{thebibliography}{99}
\bibitem{note1} Notice that, due to the structural disorder, 
the density energy Eq.~(\ref{energy}) for a SN lacks cylindrical symmetry around the $x$ axis, 
and therefore in principle a single angle $\phi$ is not enough to specify 
the application of an external field with a orientation other than the longitudinal one.
For example, the hysteresis loops obtained by applying $h$ along the $y$ and $z$ axes
(both corresponding to $\phi=90^{\circ}$) should be different from each other. 
However, these differences decrease as $N$ increases and become negligible for $N \ge 100$,
and hence the angle $\phi$ by itself is sufficient to characterize this phenomenon.
\bibitem{note2} Assuming that the magnetization is aligned with the external field, 
it is easy to verify that the dipolar field is perpendicular to $h$ for $\phi \approx 54.7^{\circ}$ .
\end{thebibliography}
\end{document}